\documentstyle[12pt]{article}
\def\1{\mbox{I\hspace{-.15em}1}}

\def\b{\begin{equation}}
\def\e{\end{equation}}
\def\bee{\begin{enumerate}}
\def\eee{\end{enumerate}}

\title{Casimir effect  on  nontrivial  topology spaces\\ in  Krein space
quantization}
\author{M. Naseri$^{1,2}$
 , S. Rouhani$^{3}$ , M.V. Takook$^{2}$\thanks{e-mail:
takook@razi.ac.ir} }
\date{\today}

\begin{document}

\maketitle {\it \centerline{$^1$ Islamic Azad University,
Kermanshah Branch, Kermanshah, IRAN}\centerline{\it $^2$
Department of Physics, Razi University, Kermanshah, IRAN}
\centerline{\it $^3$ Plasma Physics Research Centre, Islamic Azad
University,}} \centerline{\it P.O.BOX 14835-157, Tehran, IRAN}

\begin{abstract}

Casimir effect of a topologically nontrivial two-dimensional
space-time, through Krein space quantization \cite{gareta,ta4},
has been calculated. In other words, auxiliary negative norm
states have been utilized here. Presence of negative norm states
play the role of an automatic renormalization device for the
theory. The negative norm states (which do not interact with the
physical world) could be chosen in two perspective. In the first
case our method results in zero or vanishing values for energy. In
the second case, however, the result are the same as the
renormalization procedure.

\end{abstract}

\vspace{0.5cm} {\it Proposed PACS numbers}: 04.62.+v, 03.70+k,
11.10.Cd, 98.80.H \vspace{0.5cm}

\newpage

\section{Introduction}

Consideration of the negative norm states was proposed by Dirac in
1942. In 1950, Gupta applied this idea in QED. The presence of
higher derivatives in the lagrangian also led to ghosts states
with negative norms. In order to preserve the covariance principle
in the gauge theory, the auxiliary negative norms states were
utilized. In previous paper \cite{gareta}, it was shown that
consideration of the negative norm states is necessary for a fully
covariant quantization of the minimally coupled scalar field in de
Sitter space (Krein QFT). We have shown that for physical states
(positive norm states) the energy is positive, whereas, for the
negative norm states (so called un-physical states) the energy is
negative. It was also shown that the effect of these un-physical
states merely appears in the physics of the problem as a tool for
an automatic renormalization of the theory in one-loop
approximation \cite{gareta,ta4,ta3,ta,garota,ta1,ta2}.

In a previous paper \cite{khnarota}, the Casimir effect in Krein
QFT has been studied as well. Once again it is found that the
theory is automatically renormalized. This method is once again
reexamined here by analysis of Casimir effect in space-time with
nontrivial topology. The paper is organized as follows. The next
section presents a brief review of ordinary Casimir effect in
two-dimensional space-time with nontrivial topology.  Section 3 is
devoted to study of the vacuum energy density of scalar field in
two-dimensional space-time with $R \times S^1$ topology. Finally,
the results are discussed and analyzed in Section 5.

\section{Casimir energy: a review}

Consider a real scalar field $\varphi(t, x)$ defined on an
interval $0<x<a$ in an one-dimensional space with $S^1$ topology.
In this case, the boundary conditions can be written as \b
\varphi(t,0)=\varphi(t,a),\;\;\partial_{x}\varphi(t,0)=\partial_{x}\varphi(t,a).\e
The scalar field equation is  \b(\Box+m^{2})\varphi(t, x)=0 .\e
The scalar product associated with this equation is \b
(f,g)=i\int_{t=const.}dx(f^{\ast}\partial_{0}g-g\partial_{0}f^{\ast})
,\e where  $f$ and $g$ are solutions of the eq. (2). It can be
seen that the positive- and negative-frequency solutions of eq.
(2) are \b u_p(k,t,x)=\frac{e^{i k x-iwt}}{\sqrt{(2\pi)2w}}
,\;\;u_n(k,t,x)=\frac{e^{-ik x+iwt}}{\sqrt{(2\pi)2w}}. \e These
modes are orthonormalized by the following relations:
$$ (u_p(k,x,t),u_p(k',x,t))=\delta( k- k'),$$ $$
(u_n(k,x,t),u_n(k',x,t))=-\delta( k-k'),$$ \b
(u_p(k,x,t),u_n(k',x,t))=0.\e $u_p$ modes are positive norm states
and the $u_n$'s are negative norm states. By imposing the boundary
conditions (1) into (4) the positive- and negative frequency
solutions can be obtained as follows: \b
\varphi_{N}^{\pm}(t,x)=\frac{1}{\sqrt{(2 a \omega_{N})}}\exp[\pm
i(\omega_{N}t-k_{N}x)] ,\e where $$
\omega_{N}=(m^{2}+k_{N}^{2})^{\frac{1}{2}},\;\; k_{N}=\frac{2 \pi
N}{a},\;\; N=0,\pm1,\pm2,...  .$$ Now the standard quantization of
the field is performed by means of the expansion \b
\phi(t,x)=\sum_{N=-\infty}^{\infty}[\varphi^{(+)}_{N}(t,x)a_{N}+\varphi^{(-)}_{N}(t,x)a^{\dag}_{N}]
.\e The energy density operator is given by the $00$-component of
the energy-momentum tensor  \b
T_{00}=\frac{1}{2}\{(\partial_{t}\phi(t,x))^{2}+(\partial_{x}\phi(t,x))^{2}\}
.\e These the vacuum energy density of a scalar field on $S^{1}$
can be calculated as follow [15]\b
\langle0|T_{00}|0\rangle=\frac{1}{2a}\sum_{N=-\infty}^{\infty}\omega_{N}
.\e The total vacuum energy is \b
E_{0}(a,m)=\int_{0}^{a}\langle0|T_{00}|0\rangle dx
=\frac{1}{2}\sum_{N=-\infty}^{\infty}\omega_{N}=\sum_{N=0}^{\infty}\omega_{N}-\frac{m}{2}
.\e

The renormalization of this infinite quantity is performed by
subtracting the contribution of the Minkowski space \b
E_{0}(a,m)=[\sum_{N=0}^{\infty}\omega_{N}-\frac{a}{2
\pi}\int_{0}^{\infty}\omega(k)dk]-\frac{m}{2} .\e Substituting
$A=\frac{am}{2 \pi}$ and $t=\frac{ak}{2\pi}$, one can obtain \b
E_{0}(a,m)=\frac{2\pi}{a}[\sum_{N=0}^{\infty}\sqrt{A^{2}+N^{2}}
-\frac{a}{2 \pi}\int_{0}^{\infty}\sqrt{A^{2}+t^{2}}
dt]-\frac{m}{2} .\e Using the Abel-Plana formula \cite{bomomo} \b
\sum
F(N)-\int_{0}^{\infty}F(t)dt=\frac{1}{2}F(0)+i\int_{0}^{\infty}\frac{dt}{e^{2\pi
t}-1}[F(it)-F(-it)] ,\e where $F(t)=\sqrt{A^{2}+N^{2}}$, we
finally obtain, \b
E_{0}^{ren}(a,m)=-\frac{4\pi}{a}\int_{A}^{\infty}\sqrt{\frac{t^{2}-A^{2}}{e^{2\pi
t }-1}}dt=-\frac{1}{\pi
a}\int_{\mu}^{\infty}\frac{\sqrt{\xi^{2}-\mu^{2}}}{e^{\xi}-1}d\xi
.\e where $\xi=2\pi t$, $\mu=2\pi A$. In the massless case ($\mu =
0$) we have \b E_{0}^{ren}(a,0)=-\frac{1}{\pi
 a}\int^{\infty}_{0}\frac{\xi}{\exp(\xi)-1}d\xi = -\frac{\pi}{6a} .\e
 For $\mu\ll1$ it follows from (13)
 \b E_{0}^{ren}(a,m)\approx -\frac{\sqrt{\mu}}{\sqrt{2 \pi
 a}}e^{-\mu} ,\e
 i.e., the vacuum energy of the massive field is exponentially small.

\section{Vacuum energy in  Krein space quantization}

In the previous paper \cite{ta4}, we present the free field
operator in the Krein space quantization \b
\phi(t,x)=\phi_p(t,x)+\phi_n(t,x),\e where
$$ \phi_p(t,x)=\int dk [a(k)\varphi_p(k,x,t)+a^{\dag}(
k)\varphi_p^*(k,x,t)],$$ $$ \phi_n(t,x)=\int dk [b(
k)\varphi_n(k,x,t)+b^{\dag}(k)\varphi_n^*(k,x,t)],$$ and $a(k)$
and $b(k)$ are two independent operators. Creation and
annihilation operators are constrained to obey the following
commutation rules \b [a( k),a(k')]=0,\;\;[a^{\dag}( k),
a^{\dag}(k')]=0,\;\;, [a( k),a^{\dag}(k')]=\delta(k-k') ,\e \b
[b(k),b( k')]=0,\;\;[b^{\dag}( k), b^{\dag}( k')]=0,\;\;, [b(
k),b^{\dag}( k')]=-\delta( k-k') ,\e
 \b [a( k),b(k')]=0,\;\;[a^{\dag}(k), b^{\dag}( k')]=0,\;\;, [a
(k),b^{\dag}(k')]=0,\;\;[a^{\dag}( k),b(k')]=0 .\e The vacuum
state $\mid \Omega>$ is then defined by \b a^{\dag}(k)\mid
\Omega>= \mid 1_{ k}>;\;\;a(k)\mid \Omega>=0, \e \b b^{\dag}(
k)\mid \Omega>= \mid \bar1_{k}>;\;\;b( k)\mid \Omega>=0, \e \b b(
k)\mid 1_{ k}
>=0;\;\; a(k)\mid \bar1_{k} >=0, \e
where $\mid 1_{k}> $ is called a one particle state and $\mid
\bar1_{k}>$ is called a one ``unparticle state''.

We are now in a position to calculate the vacuum energy of
two-dimensional space-time with nontrivial topology in Krein
space. The field operator in Krein space is build by joining two
possible solutions of field equation, positive and negative norms.
The negative norm states, which do not interact with the physical
world could be constructed with two perspective. These
perspectives lead us to build two possible field operators.

\subsection{First perspective}

In this case the boundary conditions (1) are intrinsic properties
of the space-time. Consequently both positive  and negative energy
basis are affected by these conditions. So the scalar field
operator in such this perspective (through Krein quantization
method) can be written as: \b \phi(t,x)=\sum_{N=-\infty}^{\infty}
[\varphi^{(+)}_{N}(t,x)a_{N}+\varphi^{(-)}_{N}(t,x)a^{\dag}_{N}]+\sum_{N=-\infty}^{\infty}
[\varphi^{(-)}_{N}(t,x)b_{N}+\varphi^{(+)}_{N}(t,x)b^{\dag}_{N}],\e
where $\varphi_{N}^{(\pm)}(t,x)$ is defined in (5). The vacuum
energy density of a scalar field on $S^{1}$ can be calculated as
follows \b \langle\Omega|T_{00}^{Kre}|\Omega\rangle=\frac{1}{2
a}\sum_{N=-\infty}^{\infty} \omega_{N}-\frac{1}{2
a}\sum_{N=-\infty}^{\infty} \omega_{N}=0 .\e Therefore the vacuum
energy is automatically renormalized  and it is equal to zero. In
this perspective the structure of space-time is not affected by
the vacuum energy.

\subsection{Second perspective}

In this perspective, the gravitational field appears as an
external phenomena imposed on the structure of space-time. This is
due to the fact that the boundary conditions $(1)$ are only
imposed on positive norm states.

By imposing the above boundary conditions, the field operator in
Krein QFT can be written as follows: \b
\phi(t,x)=\sum_{N=-\infty}^{\infty}[\varphi^{(+)}_{N}(t,x)a_{N}+\varphi^{(-)}_{N}(t,x)a^{\dag}_{N}]+\int
d k [b(\vec k)u_n(k,x,t)+b^{\dag}(\vec k)u^{\ast}_n(k,x,t)]\e
where $\varphi_{N}^{(\pm)}(t,x)$ and $u_{n}(k,x,t)$ are defined in
(6) and (4) respectively. By Substituting the above field operator
in (7) and using Eqs (24),(25) and (26), one can easily obtain: \b
\langle\Omega|T_{00}^{Kre}|\Omega\rangle=\frac{1}{2
a}\sum_{N=-\infty}^{\infty} \omega_{N}-\frac{1}{2
\pi}\int_{0}^{\infty} \omega(k)dk.\e The total vacuum energy in
Krein QFT is \b
E_{0}^{Kre}(a)=\int_{0}^{a}\langle\Omega|T_{00}^{Kre}|\Omega\rangle
dx=[\sum_{N=-\infty}^{\infty}\omega_{N}-\frac{a}{2\pi}\int_{0}^{\infty}\omega(k)dk]-\frac{m}{2}.\e
It is clearly seen that once again the previous result $(10)$ has
attained. It should be noted that this energy can not be detected
locally. In the semiclassical treatment of gravitional field,
however, it affects the curvature globally through
 \b R_{\mu\nu}-\frac{1}{2}Rg_{\mu\nu}=8\pi G \langle T_{\mu\nu}\rangle,
\e {\it i.e.} it changes the structure of the space-time. In this
perspective we have the quantum instability of the space-time
topology.

\section{Conclusion }

The Casimir effect is the simplest example of interaction field.
The zero point energy of quantum fields in this case, i.e. Casimir
energy, can be alternatively calculated as an interaction
lagrangian without reference to zero point energies \cite{ja}. The
goal of the presentation of Krein space quantization is to
eliminate the singularity which appears in the interaction field.
In the present paper, Casimir energy for two-dimensional
space-time with nontrivial topology, has been calculated through
the Krein space quantization. Once again it is found that the
theory is automatically renormalized. The zero point energy of
vacuum is found to be zero if we suppose that the structure of
space-time is dependent on the matter it contains. If we take the
perspective that the structure of space-time is not independent of
matter and only boundary conditions are imposed as the external
interaction, we obtain the regular results found by other authors
but we have the quantum instability of the space-time.


\vspace{0.5cm}

\begin{thebibliography}{a}
\bibitem{gareta} J.P. Gazeau, J. Renaud, M.V. Takook, Class. Quan.
Grav, 17(2000)1415, gr-qc/9904023
\bibitem{ta4} M.V. Takook, Int. J. Mod. Phys. E, 11(2002)509,
gr-qc/0006019
\bibitem{ta3}M.V. Takook, Int. J. Mod. Phys. E, 14 (2005) 219,
gr-qc/0006052.
\bibitem{ta} M.V. Takook, Th\`ese de l'universit\'e Paris VI, 1997 {\it
Th\'eorie quantique des champs pour des syst\`emes \'el\'ementaires
``massifs'' et de ``masse nulle'' sur l'espace- temps de de Sitter.}
\bibitem{garota}  J.P. Gazeau,  S. Rouhani, M.V Takook,
{\it Linear covariant quantum gravity in de Sitter space}, in
preperation
\bibitem{ta1} M.V. Takook, Mod. Phy. lett. A, 16(2001)1691,
gr-qc/0005020
\bibitem{ta2} Takook M.V., Proceeding of the Wigsym6, 16-22 August,
1999, Istanbul, Turkey, gr-qc/0001052
\bibitem{khnarota} H. Khosravi, M. Naseri, S. Rouhani  and M.V. Takook, Phys.
Lett. B 640(2006)48, gr-qc/0604036
\bibitem{ja} R.L. Jaffe, Phys. Rev. D, 72(2005)021301,
hep-th/0503158

\end{thebibliography}
\end{document}